\documentclass[prb,twocolumn,showpacs,preprintnumbers,amsmath,amssymb,floatfix]
                                                                      {revtex4}
\usepackage{graphicx}
\usepackage{dcolumn}
\usepackage{bm}
\usepackage{multirow}

\usepackage{color}

\usepackage{subfigure}

\allowdisplaybreaks[1]

\usepackage{multirow}
\begin{document}
%%%%%%%%%%%%%%%%

\title{Electronic, magnetic and transport properties of full and 
                         half-metallic thin film Heusler alloys}

\author{Denis Comtesse$^1$}  
\author{Benjamin Geisler$^1$}
\author{Peter Entel$^1$}
\author{Peter Kratzer$^1$}
\author{L\'aszl\'o Szunyogh$^2$}
\affiliation{
  $^1$Faculty of Physics, University of Duisburg-Essen and 
      CENIDE, D-47048 Duisburg, Germany}
\affiliation{
  $^2$Budapest University of Technology and Economics, 
      Department of Theoretical Physics and Condensed Matter Research 
      Group of the Hungarian Academy of Sciences, 
      Budafoki \'ut. 8, H-1111 Budapest, Hungary}

\date{\today}

%%%%%%%%%%%%%%%%
\begin{abstract}
%%%%%%%%%%%%%%%%
The electronic and magnetic bulk properties of half-metallic Heusler alloys such as
Co$_{2}$FeSi, Co$_{2}$FeAl, Co$_{2}$MnSi and Co$_{2}$MnAl 
are investigated by means of {\em ab initio} calculations in combination with Monte 
Carlo simulations. The electronic structure is analyzed using the plane wave 
code Quantum Espresso and magnetic exchange interactions are determined
using the KKR method. From the magnetic exchange interactions the
Curie temperature is obtained via Monte Carlo simulations. In addition, electronic
transport properties of the trilayer systems consisting of two semi-infinite platinum
leads and a Heusler layer in between are obtained from the fully relativistic KKR method by
employing the Kubo-Greenwood formalism. The focus is on thermoelectric properties, namely
the Seebeck effect and its spin dependence. It turns out that already thin Heusler layers
provide highly polarized currents within the systems. This is attributed to the recovery of
half-metallicity with increasing thickness. The absence of electronic states of the spin down electrons
around the Fermi level suppresses the contribution of this spin channel to the total conductivity.
This strongly influences the thermoelectric properties of such systems and results in polarized
thermoelectric currents.
%%%%%%%%%%%%%%
\end{abstract}
%%%%%%%%%%%%%%


\maketitle

%%%%%%%%%%%%%%%%%%%%%%%%%%%%%%%%%%%%%%%
\section{Introduction}\label{sec:intro}
%%%%%%%%%%%%%%%%%%%%%%%%%%%%%%%%%%%%%%%

The question how a spin current can be generated and injected into functional 
devices is of great interest for future technological applications. Recently, 
spin dependent phenomena in the field of thermoelectrics raised a discussion 
about the interplay between spin, charge and heat currents \cite{Bauer}. For 
example, there is the possibility that a strong spin dependence of 
the Seebeck coefficient can be used to generate spin accumulation by applying 
a temperature gradient \cite{Slachter}. This spin accumulation could be used to drive spin 
currents into functional devices. In other words, one is interested in a thermally
driven spin current generator.

The ferromagnetic half-metals are promising candidates in this field 
because they exhibit a 100\% spin-polarization of the electronic
density of states (DOS) at the Fermi level. This means that there is a gap in one
of the spin channels and it should, in principle, be possible to 
extract a 100\% spin-polarized current out of these materials.
Unfortunately, the sensitive dependence of half-metallicity on details like
interfaces and interface defects have up to now hindered a simple 
generation of spin currents with high polarization from the half-metals. Therefore, theoretical
investigations of possible thermoelectric devices which can inject spin currents
of high polarization, is of great importance.

In this work, results of  \textit{ab initio} simulations of Co, Fe and Mn 
based Heusler alloys are reported. These alloys are half-metallic ferromagnets and 
have already been considered for spintronics applications (see, e.g., Ref.\,\cite{Katsnelson08}). 
Structural, electronic and magnetic properties as well as the Curie 
temperatures of the alloys are determined by combining \textit{ab initio} methods with 
Monte Carlo simulations. The Curie  temperature is of special interest 
because it is required to be sufficiently high in order to allow the design of 
devices which keep the ferromagnetic half-metallicity 
beyond room temperature. Details of the half-metallic electronic structure are discussed
in terms of the symmetry resolved density of states (DOS) within the $e_{g}$ and $t_{2g}$
representation.

In addition, {\em ab initio} calculations of thermoelectric transport
properties of Co$_2$FeAl, Co$_2$FeSi, Co$_2$MnSi and Co$_2$MnAl with Pt 
contacts are performed. A strong dependence on layer thickness and 
composition is found. Furthermore, the contributions of the two spin-channels reveal the possibilty
that spin-polarized currents can be generated by applying a thermal gradient.

The Heusler system within platinum layers can be considered as an analogue to 
the copper-cobalt multilayers investigated by Gravier \textit{et al.}. 
\cite{Gravier} The authors report interesting spin-dependent electronic and thermoelectric
properties such as the magneto-resistance and magneto-thermopower. In the present work the role of the non-magnetic
copper is taken over by platinum that introduces strong spin orbit coupling and the magnetic Co
is replaced by Co based Heusler alloys.

%%%%%%%%%%%%%%%%%
\begin{figure}[t]
	\includegraphics[width=0.79\columnwidth,clip]{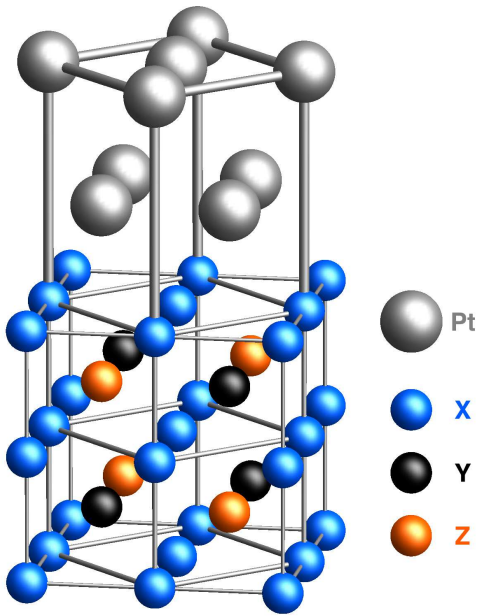}
	\caption{Schematic view of the interface between platinum 
        leads and layers. The platinum lattice is 
        rotated with respect to the Heusler lattice because the length of the 
        diagonal of the platinum lattice is comparable to the lattice 
        constants of the Heusler alloys. This ensures that the lattice 
        mismatch between both metals is between 0.06 and 0.11\,{\AA}.} 
        \label{transport_geom}
\end{figure}
%%%%%%%%%%%%

For the determination of transport properties a fully relativistic description of the electronic structure within the 
screened Korringa-Kohn-Rostoker (SKKR) \cite{SKKR} method in combination with the 
Kubo-Greenwood formalism is employed. This ensures that the spin-orbit coupling introduced
by platinum is taken into account. The relativistic spin-projection
operator introduced by Lowitzer \textit{et al.} \cite{Lowitzer} is used to evaluate the spin dependent 
contributions. This operator allows the projection of current contributions onto the two spin channels
in the relativistic framework.

The main goal of this work is to evaluate the possibility to drive a highly spin-polarized current by applying
a thermal gradient. As it is very difficult to grow large crystals of Heusler alloys with perfect L2$_{1}$
structure, a direct application of such alloys as half-metallic spin injectors is hindered because
half-metallicity is suppressed by disorder. But it is possible to grow thin layers of Heusler alloys with perfect
L2$_{1}$. Therefore, if it turns out that a thin film of half-metallic Heusler between
two leads already induces a high spin-polarization of the current, the systems under consideration here
are of special interest. It is shown throughout this paper that exactly this is possible.

The remainder of the paper is organized as follows: First, details of the 
calculations are described. Special attention is paid to the details of the transport calculations
and the modeling of the Pt-Heusler-Pt systems. Subsequently, results of calculations of the electronic structure
of the bulk Heulser system are presented followed by a discussion of the corresponding
magnetic exchange interactions and Curie temperatures. Subsequently, a detailed investigation of
transport properties and electronic structure of the Pt-Heusler-Pt systems is carried out. In the final
section the results are summarized and the conclusions drawn from the calculations are given.

%%%%%%%%%%%%%%%%%%%%%%%%%%%%%%%%%%%%%%%%%%%%%%%%%%%%%%%%%
\section{Transport theory formalism}\label{sec:Formalism}
%%%%%%%%%%%%%%%%%%%%%%%%%%%%%%%%%%%%%%%%%%%%%%%%%%%%%%%%%

The calculation of the electronic transport properties is carried out 
within the linear response formalism by employing the Kubo-Greenwood formula
%
%%%%%%%%%%%%%%%%
\begin{equation}
\sigma_{\mu\nu}(\boldsymbol{r},\boldsymbol{r}')=\frac{\hbar}{\pi N\Omega}\mathrm{Tr}\left[\hat{J}_{\mu
}G^{+}(\boldsymbol{r},\boldsymbol{r}',E)\hat{J}_{\nu}
G^{-}(\boldsymbol{r},\boldsymbol{r}',E)\right]
\end{equation}
%%%%%%%%%%%%%%
%
where $\sigma_{\mu\nu}$ is the conductivity tensor, $\hat{J}$ is the 
current operator and the $G^{+}$ and $G^{-}$ the advanced and retarded Green's function 
which is obtained from the KKR formalism and the trace is taken in the four-component Dirac-space.

The total conductance of a particular devices is expressed as
\begin{equation}
g=\int_{S_{\mathrm{R}}}\mathrm{d}S\int_{S_{\mathrm{L}}}\mathrm{d}S^{\prime}\,\hat{\boldsymbol{n}}\,\boldsymbol{\sigma}\left(\boldsymbol{r},\boldsymbol{r}^{\prime}\right)\hat{\boldsymbol{n}}^{\prime}
\end{equation}
where the $S$ and $S'$ are surfaces in the asymptotic region of the leads and
$\hat{\boldsymbol{n}}$, $\hat{\boldsymbol{n}}'$ are the corresponding normal vectors.
To distinguish between contributions of the two spin channels, the
fully relativistic spin-projection operator \cite{Lowitzer}
%
%%%%%%%%%%%%%%%%
\begin{equation}
\hat{\mathcal{P}}_{z}^{\pm} = \frac{1}{2} \left[
      1\pm\left(\beta\Sigma_{z}-\frac{\gamma_{5}\hat{p}_{z}}{mc}\right) \right]
\end{equation}
%%%%%%%%%%%%%%
%
is employed to define the spin-projected current operators $\mathcal{J}^{\pm}=\hat{\mathcal{P}}_{z}^{\pm}\hat{J}$
and the spin current operator $\mathcal{J}=(\hat{\mathcal{P}}_{z}^{+}-\hat{\mathcal{P}}_{z}^{-})\hat{J}$.
Therewith, a correlation between a current induced in one lead and the spin-polarized response current in
the other lead is determined. From here on the two spin channels are denoted by $\uparrow$ and $\downarrow$ instead
of $\pm$.

The Seebeck coefficient is evaluated using the approach of Sivan and Imry 
\cite{Sivan} who defined the moments
%
%%%%%%%%%%%%%%%%
\begin{equation}
L_{n}=-\int g\left(E\right)\left(E-\mu\right)^{n}
\left[\frac{\partial}{\partial E}f\left(E,\mu,T\right)\right]\mathrm{d}E
\end{equation}
%%%%%%%%%%%%%%
%
from which the Seebeck coefficient
%
%%%%%%%%%%%%%%%%
\begin{equation}
S=-\frac{1}{eT}\frac{L_{1}\left(\mu,T\right)}{L_{0}\left(\mu,T\right)}
\label{SeebeckCoeff}
\end{equation}
%%%%%%%%%%%%%%
%
can be calculated. Using this approach two types of spin-dependent thermoelectric quantities
can be defined. The first one is obtained by splitting the numerator of
Eq.\ (\ref{SeebeckCoeff}) into two additive contributions, which leads to the definition
%
%%%%%%%%%%%%%%%%
\begin{equation}
\tilde{S}_{\sigma}=-\frac{1}{eT}\frac{L_{1,\sigma}
  \left(\mu,T\right)}{L_{0}\left(\mu,T\right)}
  ,\hspace{0.5cm}\sigma=\uparrow,\downarrow\,.
\label{SeebeckCoeffTilde}
\end{equation}
%%%%%%%%%%%%%%
%
where the quantities $\tilde{S}$ should not be confused with the Seebeck 
coefficient of a single spin channel.

The definition in Eq.\ \eqref{SeebeckCoeffTilde} gives insight into how the two spin 
channels give additive contributions to the total Seebeck coefficient. It allows to determine
which of both channels is responsible for the major contribution.

Employing the spin current operator leads to
%
%%%%%%%%%%%%%%%%
\begin{equation}
\tilde{S}_{\mathrm{spin}}=-\frac{1}{eT}\frac{L_{1,\uparrow}
\left(\mu,T\right)-L_{1,\downarrow}
\left(\mu,T\right)}{L_{0}\left(\mu,T\right)}
\label{SpinSeebeckCoeff}
\end{equation}
%%%%%%%%%%%%%%
%

which is understood as a measure of the spin accumulation 
that is driven by the thermal gradient.

%%%%%%%%%%%%%%%%%%%%%%%%%%%%%%%%%%%%%%%%%%%%%%%%%%%%%%
\section{Computational details}\label{sec:CompDetails}
%%%%%%%%%%%%%%%%%%%%%%%%%%%%%%%%%%%%%%%%%%%%%%%%%%%%%%

The determination of structural parameters of the bulk material is 
carried out with Quantum Espresso \cite{QEspress}, where a relaxation 
of the volume is performed and tendencies for tetragonal distortions
are examined. The GGA exchange correlation functional of 
Perdew, Burke and Ernzernof (PBE)  \cite{PBE96} is used because the structural 
parameters of metals obtained with this functional are in good agreement
with experimental data.

Calculations of magnetic exchange parameters are carried out 
with the SPR-KKR package \cite{Ebert96,Ebert99}. Therefore, a scalar 
relativistic determination of the multiple scattering properties is performed using the PBE functional.
The structural parameters obtained from Quantum Espresso serve as input for the KKR calculation.
Afterwards, the magnetic exchange parameters are calculated from the scattering proberties
by employing the Lichtenstein formalism \cite{Lichtenstein84,Lichtenstein86}.

The exchange parameters are subsequently used in Monte Carlo 
(MC) simulations of the classical Heisenberg model to determine the Curie temperature.
The exchange interactions are cut off after three lattice constants. This is sufficient
because exchange parameters of larger distance are too small to be relevant
for the theoretical estimation of the Curie temperature. The simulation box 
used within the MC simulations includes $10\times 10\times 10$ Heusler unit 
cells.

Calculations of  transport properties are carried out using the fully relativistic screened KKR formalism. \cite{SKKR} 
The linear response formalism in the formulation of Baranger and Stone \cite{BarangerStone} 
as first implemented by Mavropoulos \textit{et al.} \cite{Phivos} is used. In order to 
ensure an accurate determination of the transport properties, 
more than 90,000 $k$-points within the irreducible wedge of the 
two-dimensional Brillouin zone are used. The imaginary part is set 
to $0.0001$\,Ry. The energy grid for the calculation of the Seebeck coefficient 
is $0.001$\,Ry.

%%%%%%%%%%%%%%%%%%%%%%%%%%%%%%%%%%%%%%%%%%%
\section{Details of the transport geometry}
\label{sec:DetailsOfTheTransportGeometry}
%%%%%%%%%%%%%%%%%%%%%%%%%%%%%%%%%%%%%%%%%%%

Details of the geometry used in the transport calculations 
are shown in Fig.\ \ref{transport_geom}. The layer distance at the interface 
between platinum and the Heusler system is the averaged 
value of the layer distance in platinum and that in the Heusler. 
No lattice relaxations are included.

Within the SKKR method the system is assumed to be translational invariant in the 
$x$- and $y$-direction and in the $z$-direction the system is terminated on both sides
by two semi-infinite leads. Due to the two dimensional translational invariance 
a two-dimensional lattice constant $a_{2d}$ has to be defined by
%
%%%%%%%%%%%%%%%%
\begin{equation}
a_{2d}=\sqrt{2}a_{bcc}
\end{equation}
%%%%%%%%%%%%%%
%
where $a_{bcc}$ is the three-dimensional lattice constant of the underlying 
bcc lattice of the Heusler part of the system. Hence, the distance between two
 subsequent monolayers is given by
%
%%%%%%%%%%%%%%%%
\begin{equation}
d=\frac{a_{3d}}{2}=\frac{\sqrt{2}}{4}a_{2d}\,.
\end{equation}
%%%%%%%%%%%%%%
%
The atomic positions are defined for the following basis vectors using a frame 
of reference that is rotated by $45^{\circ}$ with respect to that of Fig.\ 1:
%
%%%%%%%%%%%%%
\begin{align}
\mathbf{a}_{1}  & =a_{2d}(1,0,0)\\
\mathbf{a}_{2}  & =a_{2d}(0,1,0)\nonumber\\
\mathbf{a}_{3}  & =a_{2d}(0,0,1)\nonumber\,.
\end{align}
%%%%%%%%%%%
%
The platinum and the Heusler lattices are rotated by $45^{\circ}$ with respect 
to each other. This ensures the smallest possible lattice 
mismatch between the two structures. For example, the lattice constant of 
the Heusler cell of Co$_{2}$FeSi is $2.81$\,{\AA}
(which is half of the lattice constant of the 16 atoms cell) and
the lattice constant of Pt is $3.92\,\mathrm{\mathring{A}}$. This is
of the same order as the diagonal of the Heusler structure which is  
$3.97\,\mathrm{\mathring{A}}$ and therefore close to the Pt lattice constant.
The lattice mismatch is between 0.06 to 0.11\,{\AA} depending on the particular
Heusler system.

For the calculations it is assumed that the lattice constant of Pt is the 
same as that of the two-dimensional lattice constant of the Heusler
system:
%
%%%%%%%%%%%%%%%%
\begin{equation}
a_{3\mathrm{d,Pt}}=a_{2\mathrm{d,Heusler}}
\end{equation}
%%%%%%%%%%%%%%
%
The atomic volume of the Heusler is $a_{bcc}^{3}/2=a_{2d}^{3}
/4\sqrt{2}$, hence, the (average) Wigner-Seitz radius is calculated by
%
%%%%%%%%%%%%%%%%
\begin{eqnarray}
\frac{4\pi}{3}R_{ws}^{3}&=&\frac{a_{2d}^{3}}{4\sqrt{2}}\Rightarrow R_{ws}
^{H}=\frac{1}{4\sqrt[6]{32}}\left(  \frac{3}{4\pi}\right)  
         ^{1/3}a_{2d}
\nonumber\\
&\simeq&0.138\;a_{2d} .
\end{eqnarray}
%%%%%%%%%%%%%%
%
In the fcc Pt lattice the atomic volume is 
$a_{fcc}^{3}/4=a_{2d}^{3}/4=2a_{bcc}^{3}=a_{2d}^{3}/\sqrt{2}$,
thus,
%
%%%%%%%%%%%%%%%%
\begin{eqnarray}
\frac{4\pi}{3}R_{ws}^{3}&=&\frac{a_{2d}^{3}}{4}\Rightarrow R_{ws}
^{Pt}=\frac{1}{\sqrt[3]{4}}\left(  \frac{3}{4\pi}\right)  ^{1/3}a_{2d}\nonumber\\
&\simeq&0.391\;a_{2d}
\end{eqnarray}
%%%%%%%%%%%%%%
%
Four layers of Pt are considered in each interface region to join 
smoothly to the two semi-infinite bulk regions. The Heusler layer is always terminated by a
Co monolayer on both sides. Therefore, the interface between Pt and the Heusler
system is always metallic. The distance between Pt and Co monolayer in the interface
is taken to be the average of the Pt and the Heusler interlayer distances, i.e.\ 
$(1/2+\sqrt{2}/4)/2=(2+\sqrt{2})/8$. 
In this way, the atomic radii for all Pt atoms can be taken as 
$R_{ws}^{\mathrm{Pt}}$ and all the atoms in the Heusler alloy can have an atomic radius 
of $R_{ws}^{\mathrm{H}}$ (H=Heusler) as above, irrespective of their atomic positions.

The assumption of such an interface structure between platinum and Heusler system
turns out to be reasonable because structural relaxation calculations do not reveal
strong changes. Instead the relaxations obtained from such configurations are very
small. Therefore, it is concluded that the description of the interface chosen here serves
as a good approximation.

%%%%%%%%%%%%%%%%%%%%%%%%%%%%%%%%%%%%
\section{Results}\label{sec:results}
%%%%%%%%%%%%%%%%%%%%%%%%%%%%%%%%%%%%

At first a discussion of electronic and magnetic properties of bulk 
Co$_{2}$FeAl, Co$_{2}$FeSi, Co$_{2}$MnAl and Co$_{2}$MnSi is presented. 
Table \ref{tab:sumar_table} summarizes structural properties, 
magnetic moments and Curie temperatures of all systems under consideration.
The magnetic moments are close to the expected integer values \cite{Dederichs02}
except for Co$_{2}$FeSi where a significant deviation from the expected 5$\mu_{\mathrm{B}}$ is found.
This shows that the \textit{ab initio} description of this particular 
system is deficient compared to the description of the other half-metallic
Heusler systems. This problem has been discussed extensively in literature (see Ref.\,\cite{Thoene09,Kandpal06,Wurmehl72,Balke06}).

%%%%%%%%%%%%%%
\begin{table*}
\centering
\begin{tabular}{l c c c c c c} \hline\hline \\[-0.3cm]
  Composition     & Magnetic order & $a$ (\r{A})      & $c/a$    &$\mu_{\mathrm{tot}}\left(\mu_{\mathrm{B}}\right)$      & $T_{\mathrm{C}} (K)$   & $T_{\mathrm{C,exp}} (K)$  \\[0.05cm] \hline\\[-0.3cm]
  Co$_{2}$FeAl        & FM                     & 5.70                   & 1     & 4.99               & 1050                     & -                                             \\ [0.1cm]
  Co$_{2}$FeSi        & FM                     & 5.63                   & 1     & 5.57              & 750                       & 1100                                      \\ [0.1cm]
  Co$_{2}$MnAl        & FM                    & 5.70                   & 1     & 4.03             & 480                        & 697                                        \\ [0.1cm]
  Co$_{2}$MnSi        & FM                    & 5.63                   & 1     & 5.00             & 755                        & 985                                       \\ [0.05cm]
 \hline\hline
	\end{tabular}
  	\caption{Lattice parameters and magnetic order obtained from 
Quantum Espresso (GGA) calculations. In addition, the critical temperatures obtained 
with MC simulations are listed. The experimental value of the critical 
temperature of Co$_{2}$FeAl is not known because it is located close to a structural 
transition.} 
\label{tab:sumar_table}
\end{table*}
%%%%%%%%%%%%

In the following a discussion of the electronic structure and the magnetic properties of the
bulk systems is given. Afterwards the transport properties and in particular the Seebeck coefficient
of the trilayer systems are presented. The main features found in the calculation of the Seebeck
coefficient are subsequently discussion on the basis of the layer resolved DOS of the trilayers.
%%%%%%%%%%%%%%%%%%%%%%%%%%%%%%%%%%%%%%%%%%%%%%%%%%%%%%%%%%%%%%%%
\subsection{Electronic structure}\label{sec:Electronicstructure}
%%%%%%%%%%%%%%%%%%%%%%%%%%%%%%%%%%%%%%%%%%%%%%%%%%%%%%%%%%%%%%%%

\begin{figure}[t]
	\includegraphics[width=0.92\columnwidth,clip]{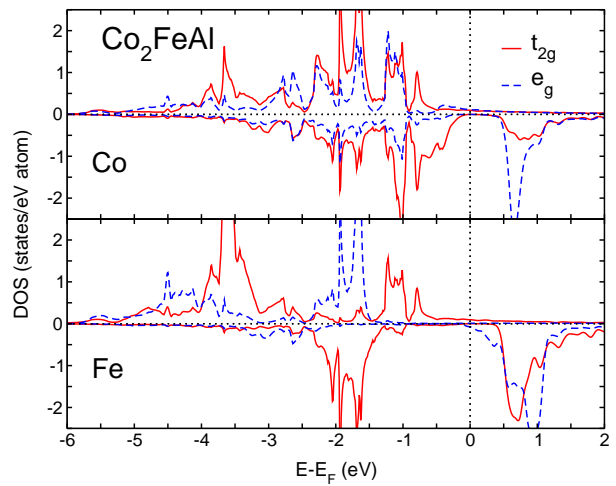}
	\caption{Symmetry resolved electronic density of states of cobalt 
and iron in Co$_{2}$FeAl. This result shows that there is a gap in the 
density in the minority spin channel. The Fermi energy lies at the upper 
edge of this gap.} 
\label{DOS_lm_Co2FeAl}
\end{figure}

Figure \ref{DOS_lm_Co2FeAl} shows the symmetry resolved electronic density 
of states of cobalt and iron of bulk Co$_{2}$FeAl. The DOS shows that 
Co$_{2}$FeAl is half-metallic because there is a pronounced gap in the 
minority spin channel and the Fermi energy is located exactly at the upper 
edge of the gap. The total magnetic moment of $4.99\mu_{\mathrm{B}}$ shows 
that the calculations nicely reproduce the expected integer value, 
revealing only a very small deviation.

%%%%%%%%%%%%%%%%%
%%%%%%%%%%%%

\begin{figure}[t]
	\includegraphics[width=0.92\columnwidth,clip]{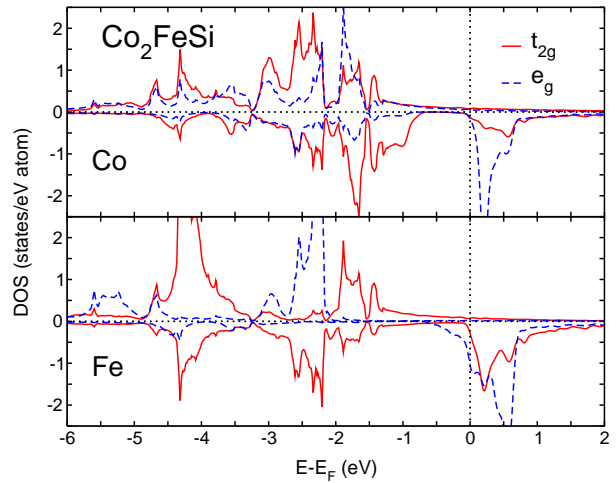}
	\caption{Symmetry resolved electronic density of states of cobalt 
and iron in Co$_{2}$FeSi. However, this calculation does not reproduce the 
expected half-metallic behavior. Although, there is a gap in the DOS of the 
minority spin channel, the Fermi energy is not located within the gap.} 
\label{DOS_lm_Co2FeSi}
\end{figure}

Figure \ref{DOS_lm_Co2FeSi} shows the symmetry resolved electronic DOS of 
cobalt and iron in Co$_{2}$FeSi. There is also a gap in the minority spin 
channel but the Fermi energy is not located within this gap. This may be attributed
to some kind of a breakdown of standard GGA for the description of the electron structure
of half-metallic Co$_{2}$FeSi. This has been discussed in more detail in literature (see e.g. Ref.\ 
\cite{FecherCoFeSi}). It is argued that strong correlations of the 
$d$-electrons are responsible for this breakdown and in addition it is 
demonstrated that this lack can be cured by employing the LDA+U method.
The DOS in Fig.\,\ref{DOS_lm_Co2FeSi} showes that in particular the description of the $e_{g}$ states of iron 
is deficient because some of the states that should belong 
to the conduction band leak into the half-metallic gap. The total magnetic moment of $5.57\,\mu_{\mathrm{B}}$ is 
by more than $0.4\,\mu_{\mathrm{B}}$ smaller than the expected integer 
value of $6.0\,\mu_{\mathrm{B}}$. This problem has to be kept in mind for the
interpretation of the later results.

%%%%%%%%%%%%%%%%%
%%%%%%%%%%%%

\begin{figure}[t]
	\includegraphics[width=0.92\columnwidth,clip]{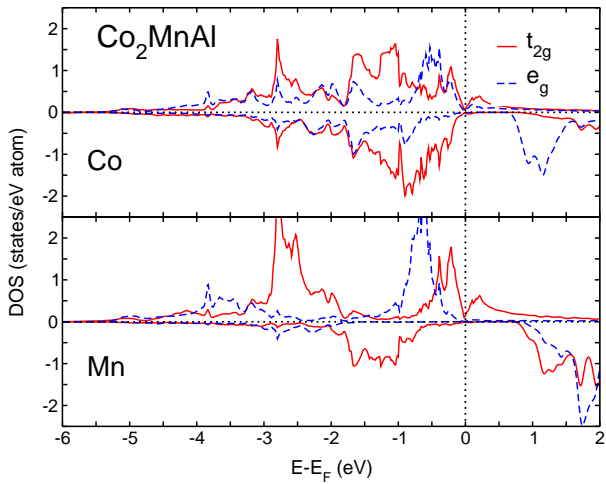}
	\caption{Symmetry resolved electronic density of states of cobalt 
and iron in Co$_{2}$MnAl. There is a pronounced gap in the minority spin 
channel and the Fermi energy is located at the lower edge of this gap.} 
\label{DOS_lm_Co2MnAl}
\end{figure}

In Fig. \ref{DOS_lm_Co2MnAl} the symmetry resolved electronic DOS of 
Co and Mn in Co$_{2}$MnAl is shown. There is an obvious gap in 
the minority spin channel. Therefore, the GGA description is sufficient in this
case and clearly captures the half-metallic properties of Co$_{2}$MnAl. The 
Fermi energy is located at the lower edge of the gap in contrast to 
the case of Co$_{2}$FeAl where the Fermi energy is located at the upper edge. 
In addition, the magnetic moments is close to the expected integer value.

%%%%%%%%%%%%%%%%%
%%%%%%%%%%%%

\begin{figure}[t]
	\includegraphics[width=0.92\columnwidth,clip]{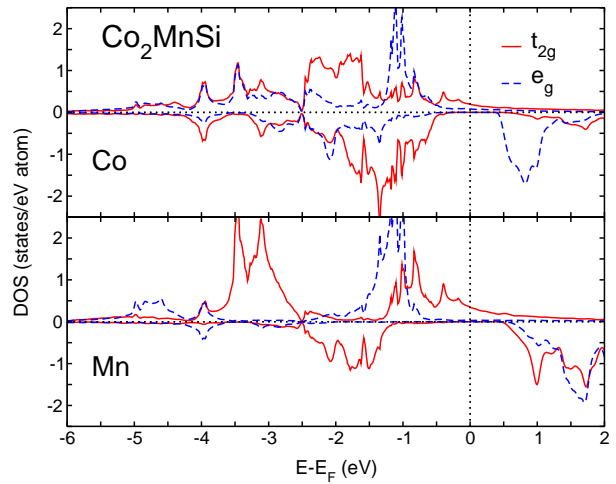}
	\caption{Symmetry resolved electronic density of states of cobalt 
and iron in Co$_{2}$MnAl. There is a pronounced gap in the minority spin 
channel and the Fermi energy is located in the middle of the gap.} 
\label{DOS_lm_Co2MnSi}
\end{figure}

The same DOS is shown in Fig. \ref{DOS_lm_Co2MnSi} for the case of
Co$_{2}$MnSi. There is a gap in the minority 
channel and the Fermi energy is located in the middle of this gap. 
Therefore, the half-metallicity of Co$_{2}$MnSi is very robust against 
interfering effects such as increasing temperature.

%%%%%%%%%%%%%%%%%
%%%%%%%%%%%%

In summary, the electronic structure of all Heusler alloys except 
Co$_{2}$FeSi is in nice agreement with the expected half-metallic structure.
Only in Co$_{2}$MnSi, the Fermi energy is located in the middle of the half-metallic
gap. This implies that the conductance of the Pt-Co$_{2}$MnSi-Pt system in an
energy interval around the Fermi level ($E_{\mathrm{F}}$) is manly given be spin up electrons.
Some contributions from the spin down channel may remain because of tunnel effect
transmission through the area where no spin down states are present. Therefore,
the Seebeck coefficient which is calculated from the conductance around
$E_{\mathrm{F}}$ must show a pronounced contribution from the
spin up and a minor one from the spin down channel.

%%%%%%%%%%%%%%%%%%%%%%%%%%%%%%%%%%%%%%%%%%%%%%%%%%%%%%%%%%%%%%
\subsection{Magnetic properties}\label{sec:Magneticproperties}
%%%%%%%%%%%%%%%%%%%%%%%%%%%%%%%%%%%%%%%%%%%%%%%%%%%%%%%%%%%%%%

%%%%%%%%%%%%%%%%%
\begin{figure}[t]
	\includegraphics[width=0.92\columnwidth,clip]{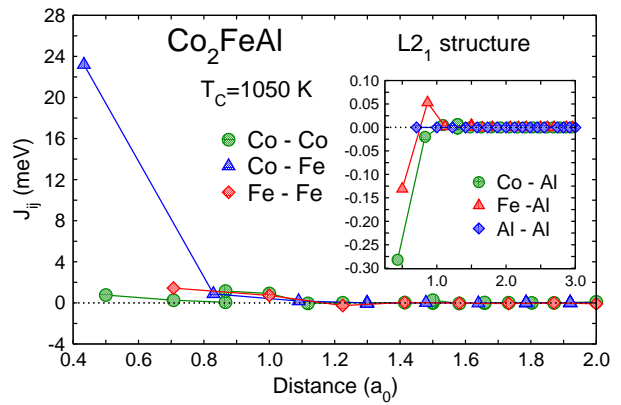}
	\caption{The magnetic exchange parameters and the critical 
        temperature of Co$_{2}$FeAl. The strongest contributions to the 
        magnetic exchange interactions are those of the cobalt-iron pairs. 
        All other interactions are small.} 
\label{Jij_Co2FeAl}
\end{figure}
%%%%%%%%%%%%

In this subsection the magnetic exchange parameters and critical 
temperatures of the Heusler bulk systems are discussed in order to compare 
between theoretical predictions of $T_{\mathrm{C}}$ and experimentally 
observed values. By investigating the calculated magnetic exchange interactions 
it becomes obvious that the magnetic structure is dominated mainly by the interaction between
cobalt and iron or between cobalt and manganese.

In Fig. \ref{Jij_Co2FeAl} the exchange interactions of Co$_{2}$FeAl are 
shown. Obviously, the most relevant interaction is the one between nearest 
neighbor cobalt and iron atoms. All other interactions are much smaller and 
almost negligible. Unfortunately, the predicted critical temperature, which is
about 1050\,K cannot be compared to experimental values because a 
structural transition from the ordered L2$_{1}$ to a disordered B2 structure
interferes with the temperature region where the magnetic transition is located.
This prevents the identification of the exact temperature of the magnetic transition.
Nevertheless, the theoretical prediction meets the expected range of the Curie
temperature.

%%%%%%%%%%%%%%%%%
\begin{figure}[t]
	\includegraphics[width=0.92\columnwidth,clip]{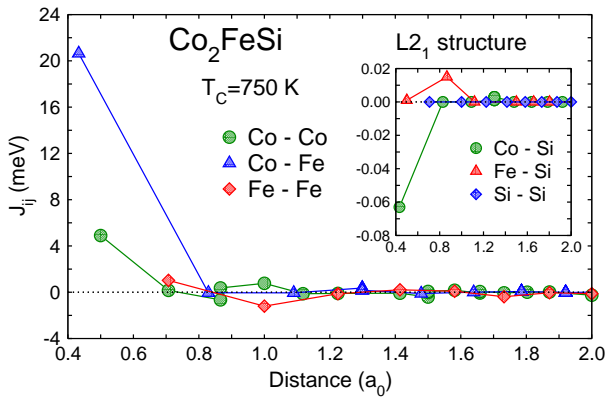}
	\caption{The magnetic exchange parameters and the critical 
temperature of Co$_{2}$FeSi. The critical temperature is much smaller than 
the experimental one. This is again due to the insufficient description of the 
electronic structure of this system by the density theory functional.} 
\label{Jij_Co2FeSi}
\end{figure}
%%%%%%%%%%%%

Figure \ref{Jij_Co2FeSi} shows the magnetic exchange interactions and the 
Curie temperature of Co$_{2}$FeSi. The exchange interactions are 
qualitatively and quantitatively comparable to those of Co$_{2}$FeAl. 
The most obvious difference is that the cobalt-cobalt interactions are 
stronger in the latter case. The Curie temperature is much smaller than 
the experimental one. This can be attributed to the incorrect description of the 
electronic structure within the GGA. As discussed in subsection 
\ref{sec:Electronicstructure} an explicit inclusion of correlation effects 
would improve the description of the electronic structure. Therefore,
it is expected that correlations will also improve the description of 
the exchange parameters and the theoretical prediction of the Curie 
temperature.

%%%%%%%%%%%%%%%%%
\begin{figure}[t]
	\includegraphics[width=0.92\columnwidth,clip]{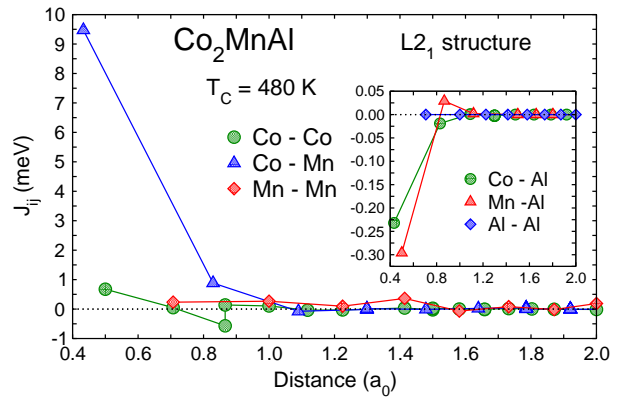}
	\caption{The magnetic exchange parameters and critical temperature 
of Co$_{2}$MnAl.} 
\label{Jij_Co2MnAl}
\end{figure}
%%%%%%%%%%%%

In the following the cobalt-manganese based Heuslers are discussed.
Although, manganese is known to introduce strong antiferromagnetic interactions 
in  the system Co$_{2}$MnAl and Co$_{2}$MnSi are clearly ferromagnets.
This can be understood by considering that the antiferromagnetism betwee Mn pairs
has a complicated distance dependence and therefore only pair which are separated
by certain distances interact antiferromagnetically. \cite{Sasioglu}

In Fig. \ref{Jij_Co2MnAl} the exchange interactions and the Curie 
temperature of Co$_{2}$MnAl is shown. Obviously, the interaction between cobalt and 
manganese is now the strongest contribution, although it is by more than a 
factor of two smaller compared to the Co-Fe interaction in cobalt-iron based Heusler systems. 
The strong Co-Mn interaction is somewhat surprising because in binary 
CoMn the exchange interaction between nearest neighbor pairs is usually 
small, negative and therefore induces antiferromagnetic coupling.

Although, the half-metallic nature of the DOS is nicely reproduced,
the critical temperature deviates significantly from experiment. This is 
due to the use of LDA in the calculation of the exchange parameters. 
There is a general trend that LDA tends to underestimate magnetic 
exchange parameters and therefore leads to smaller critical temperatures 
in the MC simulations.

%%%%%%%%%%%%%%%%%
\begin{figure}[t]
	\includegraphics[width=0.92\columnwidth,clip]{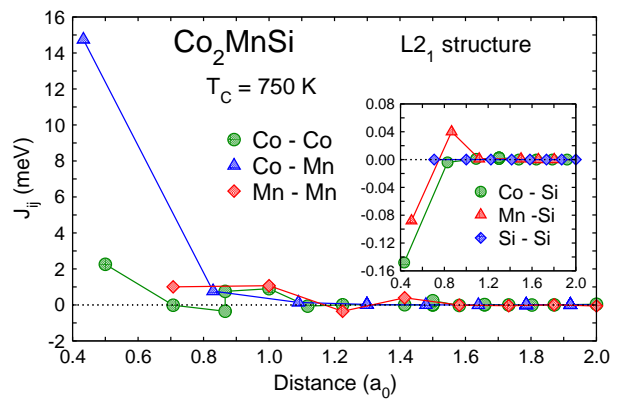}
	\caption{Magnetic exchange parameters and the critical temperature 
of Co$_{2}$MnSi.} 
\label{Jij_Co2MnSi}
\end{figure}
%%%%%%%%%%%%

Figure \ref{Jij_Co2MnSi} shows exchange parameters and critical temperature 
of Co$_{2}$MnSi. Qualitatively, the interactions are again very similar to 
those of Co$_{2}$MnAl. Here the interactions are stronger
and therefore the critical temperature is higher compared to Co$_{2}$MnAl. As the deviation 
of the critical temperature from the experimental value is of the same 
order as for Co$_{2}$MnAl, one may conclude that the general trend of the 
critical temperature is nicely reproduced and the deviation are indeed
only due to the systematic errors introduced by the LDA.

%%%%%%%%%%%%%%%%%%%%%%%%%%%%%%%%%%%%%%%%%%%%%%%%%%%%%%%%%%%%%%%
\subsection{Transport properties}\label{sec:Transportproperies}
%%%%%%%%%%%%%%%%%%%%%%%%%%%%%%%%%%%%%%%%%%%%%%%%%%%%%%%%%%%%%%%

In this subsection transport properties, in particular the Seebeck 
coefficient and its spin dependence of Pt-Heusler-Pt system are discussed. For every type of
Heusler, systems containing five monolayers are compared to systems 
containing nine monolayers of Heusler. The number of monolayers is  chosen in a way 
that the Pt-Heusler interface is purely metallic in the sense 
that the first Heusler monolayer on both sides contains only cobalt.

%%%%%%%%%%%%%%%%%
\begin{figure}[t]
	\includegraphics[width=0.93\columnwidth,clip]{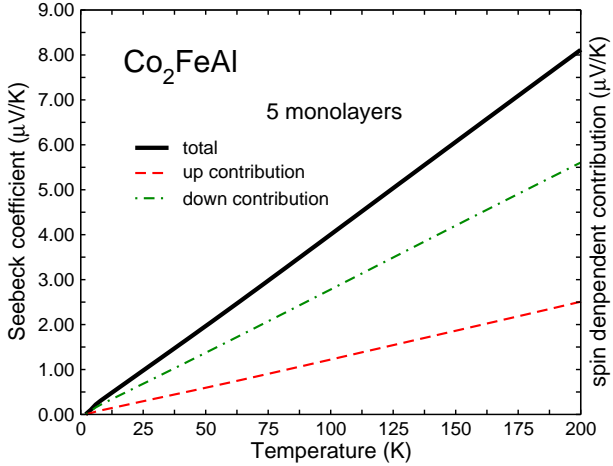}
	\caption{Seebeck coefficient of the Pt-Co$_{2}$F<eAl-Pt system with 
five Heusler monolayers between platinum leads. The black solid line 
represents the temperature dependence of the total Seebeck coefficient 
and the dashed red and green lines are the (additive) contributions from the 
spin channels to the total Seebeck coefficient.}
\label{see_Co2FeAl}
\end{figure}
%%%%%%%%%%%%

%%%%%%%%%%%%%%%%%
\begin{figure}[t]
	\includegraphics[width=0.93\columnwidth,clip]{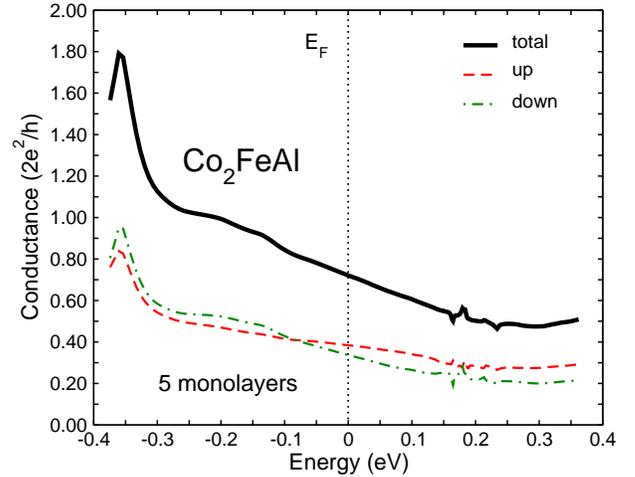}
	\caption{Energy dependence near the Fermi energy of the conductance 
of the Pt-Co$_{2}$FeAl-Pt system with five monolayers. The black line 
denotes the total conductance whereas the green and red denote the 
contributions from the two spin channels.} 
\label{conduct_Co2FeAl}
\end{figure}
%%%%%%%%%%%%

Figure \ref{see_Co2FeAl} shows the calculated temperature dependence
of the Seebeck coefficient of Pt-Co$_{2}$FeAl-Pt with five Heusler
monolayers. The Seebeck coefficient increases linearily with 
temperature. The contribution of the spin up channel is almost by a factor of 
two larger than the one of the spin down channel. In Fig.\ 
\ref{conduct_Co2FeAl} the energy dependence of the conductivity is shown. 
Around the Fermi energy the energy dependence is linear which leads to
the linear increase of the Seebeck coefficient with increasing temperature. 
For energies farther away from the Fermi level, the energy dependence of 
the conductance becomes more structured. But as these structures are 
close the the borders of the energy integration and are 
less weighted by the derivative of the Fermi function, their resulting 
contribution is only very small. Therefore the temperature dependence of
the Seebeck coefficient is still linear.

%%%%%%%%%%%%%%%%%
\begin{figure}[t]
	\includegraphics[width=0.93\columnwidth,clip]{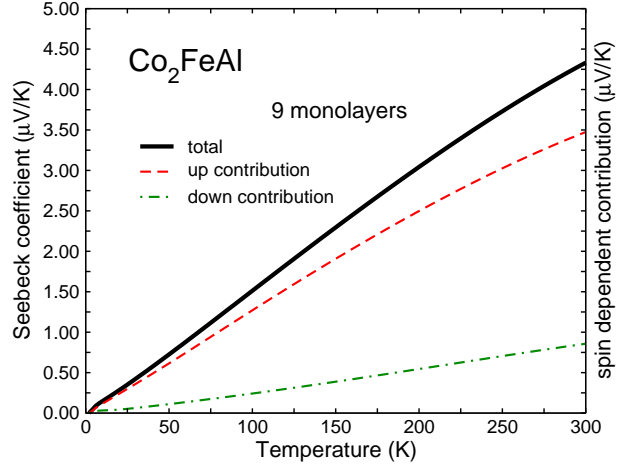}
	\caption{Calculated Seebeck coefficient of Pt-Co$_{2}$FeAl-Pt 
with nine Heusler monolayers between the platinum leads.}
\label{see_Co2FeAl2}
\end{figure}
%%%%%%%%%%%%

%%%%%%%%%%%%%%%%%
\begin{figure}[t]
	\includegraphics[width=0.93\columnwidth,clip]{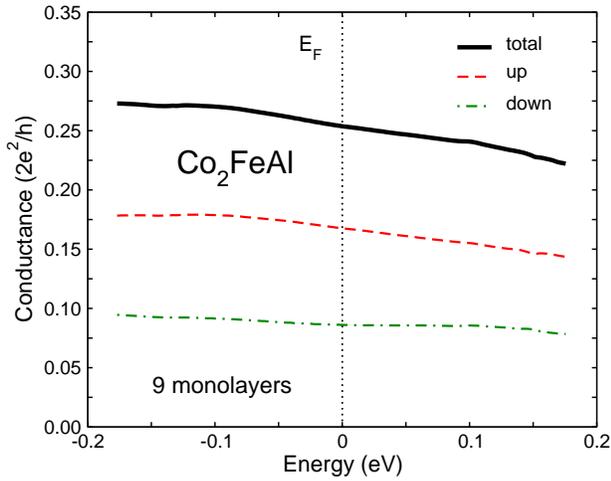}
	\caption{Energy dependence around the Fermi energy of the conductance 
of Pt-Co$_{2}$FeAl-Pt with nine monolayers.}
\label{conduct_Co2FeAl2}
\end{figure}
%%%%%%%%%%%%

As shown in Fig. \ref{see_Co2FeAl2}, the Seebeck coefficients of a system 
containing nine monolayers of Co$_{2}$FeAl is smaller compared to the system 
with five monolayers. In addition, there is a much larger difference between 
the contributions of the two spin channels to the total Seebeck coefficient. 
This can again be explained by regarding the energy dependence of the 
conductivity where the spin down channel has a slope comparable to that 
of the total conductivity and the energy dependence of the spin up channel 
is almost flat. A flat energy dependence leads to small Seebeck coefficients
because the slope of this dependence determines the size of the Seebeck
coefficient.

The first conclusion that can be drawn from the results for 
Pt-Co$_{2}$FeAl-Pt is that the Seebeck coefficient of the layered system 
depends strongly on the thickness of the layer. This basically stems from 
the different shape of the energy dependence of the conductivity. The fact 
that conductivities are very sensitive to the thickness of layers is 
well known and described e.g. in Ref.\,\cite{Phivos}.

%%%%%%%%%%%%%%%%%
\begin{figure}[t]
	\includegraphics[width=0.93\columnwidth,clip]{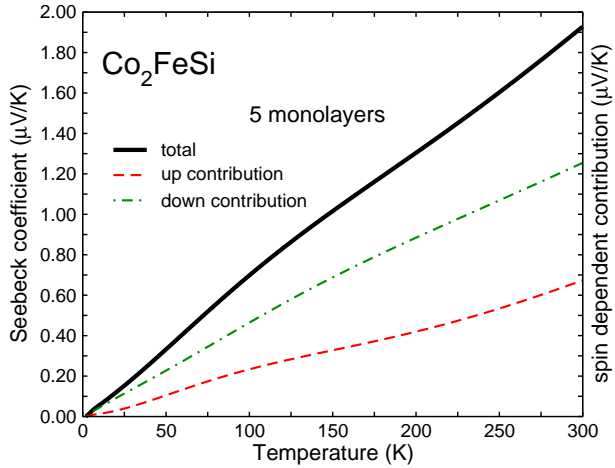}
	\caption{Seebeck coefficient of the Pt-Co$_{2}$FeSi-Pt system with 
five monolayers of Heusler between platinum leads.} 
\label{see_Co2FeSi}
\end{figure}
%%%%%%%%%%%%

%%%%%%%%%%%%%%%%%
\begin{figure}[t]
	\includegraphics[width=0.93\columnwidth,clip]{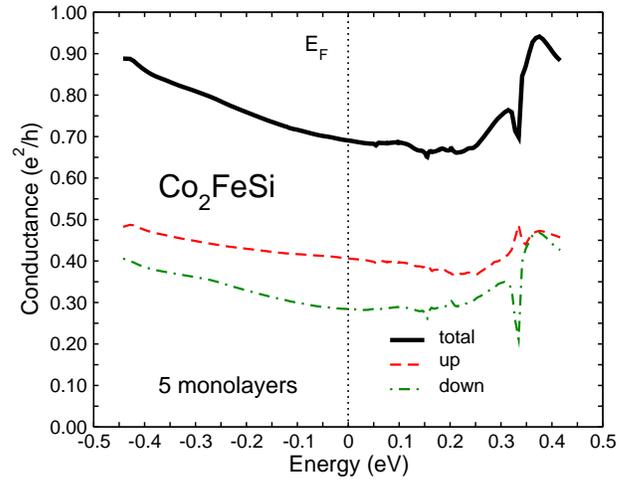}
	\caption{Energy dependence of the conductance of the 
Pt-Co$_{2}$FeSi-Pt system with nine monolayers of Heusler around the 
Fermi energy.} 
\label{conduct_Co2FeSi}
\end{figure}
%%%%%%%%%%%%

Now, the Pt-Co$_{2}$FeSi-Pt systems are discussed. This gives insight into 
how far compositional changes affect the Seebeck coefficient. In particular, the replacement
of Al by Si introduced one more valence electron. It should also be kept in mind Al is a metal
whereas Si is a semi-conductor.

Regarding Fig.\ \ref{see_Co2FeSi} which shows the temperature dependence
of the Seebeck coefficient of a system containing five monolayers of Co$_{2}$FeSi, 
it is immediately noticed that the Seebeck coefficient is by more than a factor 
of three smaller compared to the Pt-Co$_{2}$FeAl-Pt system with five monolayers.
In addition, small deviations from the linear behavior of the Seebeck coefficient are found
in this system.

The energy dependence of the conductance shown in Fig.\ 
\ref{conduct_Co2FeSi} is almost linear around $E_{\mathrm{F}}$ but reveals
a distinct structure above the Fermi energy and is still 
quite flat below. There is a pronounced peak above $E_{\mathrm{F}}$ which is
connected to numerical inaccuracies that can occur if strong changes of
the electronic structure appear during the energy sampling. In such cases the $k$-point mesh used within
the calculation can be commensurate with important features in the two dimensional
Brillouin zone at a certain energy but can miss some features at another energy.
As such structures do not affect the calculation of the Seebeck coefficient
the enormous numerical effort which is required to cure this lack
is not useful.

%%%%%%%%%%%%%%%%%
\begin{figure}[t]
	\includegraphics[width=0.93\columnwidth,clip]{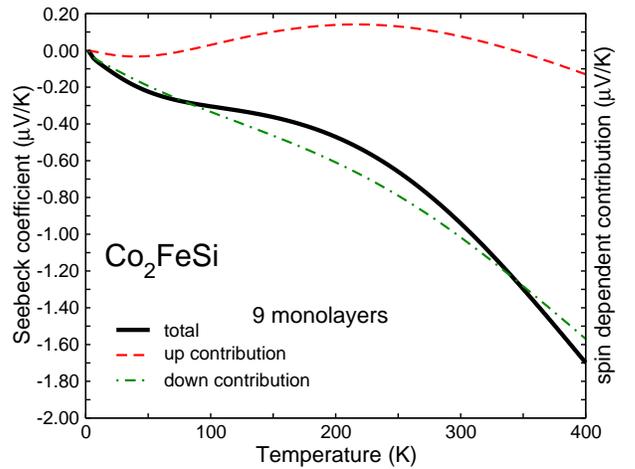}
	\caption{Seebeck coefficient of the Pt-Co$_{2}$FeSi-Pt system with 
five monolayers of Heusler between the platinum leads.} \label{see_Co2FeSi2}
\end{figure}
%%%%%%%%%%%%

%%%%%%%%%%%%%%%%%
\begin{figure}[t]
	\includegraphics[width=0.93\columnwidth,clip]{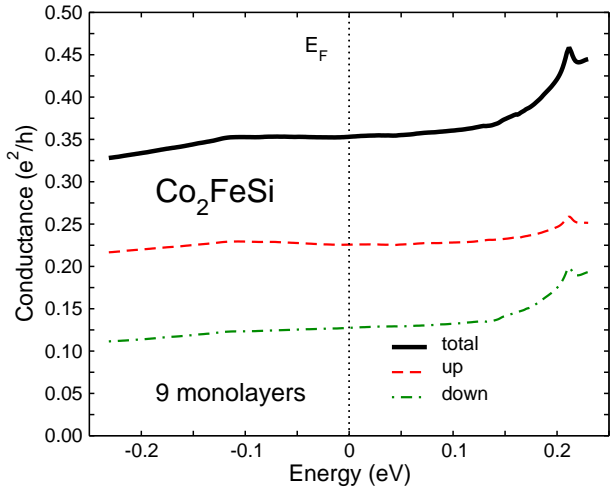}
	\caption{Energy dependence of the conductance of Pt-Co$_{2}$FeAl-Pt 
with nine monolayers of Heusler around the Fermi energy.} \label{conduct_Co2FeSi2}
\end{figure}
%%%%%%%%%%%%

Turning to the Pt-Co$_{2}$FeSi-Pt system with nine monolayers, a 
strong change occurs in comparison to the system with five monolayers. 
The Seebeck coefficient of this system is negative and the evolution with 
temperature is also not linear as for the systems with five monolayers. In addition,
the two additive contributions to the total Seebeck 
coefficient have opposite signs. The contribution of the spin down channel 
which gives the strongest contribution, is negative whereas the 
contribution of the spin up channel is small and positive. Therefore, the resulting total 
Seebeck coefficient is very small. This can be understood by
considering the energy dependence of the conductivity around the Fermi level
which is almost flat and therefore gives only small contributions to the 
Seebeck coefficient. This behavior changes around 0.1 eV away from the 
Fermi energy when more structure comes into play. Due to this the negative slope
of the Seebeck coefficient increases for temperatures above 300K.

%%%%%%%%%%%%%%%%%
\begin{figure}[t]
	\includegraphics[width=0.93\columnwidth,clip]{./see_Co2MnAl.eps}
	\caption{Seebeck coefficient of the Pt-Co$_{2}$MnAl-Pt system with 
five monolayers of Heusler between the platinum leads.} \label{see_Co2MnAl}
\end{figure}
%%%%%%%%%%%%

%%%%%%%%%%%%%%%%%
\begin{figure}[t]
	\includegraphics[width=0.93\columnwidth,clip]{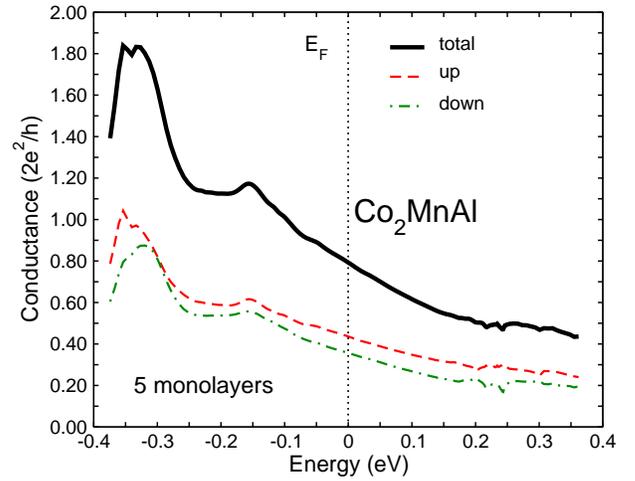}
	\caption{Energy dependence of the conductance of the 
Pt-Co$_{2}$MnAl-Pt system with nine monolayers of Heusler around the 
Fermi energy.} \label{conduct_Co2MnAl}
\end{figure}
%%%%%%%%%%%%

In a next step the Y-component of the Heusler compound is changed from iron to manganese.
This results in the occurrence of a larger Seebeck coefficient for systems containing five 
layers of Co$_{2}$MnAl. Here, the additive contributions of the spin 
channels to the total Seebeck coefficient are almost of the same size and exhibit a comparable structure. 
This energy dependence of the conductivity of this system is shown in
Fig.\ \ref{conduct_Co2MnAl}. It shows a strong slope in the total 
conductivity and both spin dependent contributions. The strong 
structure around 0.3 eV below the Fermi energy has no influence on the 
temperature dependence of the Seebeck coefficient because it is at 
too high energies to give a significant contribution to the temperature 
region considered here. Interestingly, the shape of the energy dependence of the
conductance of the Pt-Co$_{2}$FeAl-Pt system with five monolayers
is comparable to that of the Pt-Co$_{2}$MnAl-Pt with the same number of layers.

%%%%%%%%%%%%%%%%%
\begin{figure}[t]
	\includegraphics[width=0.93\columnwidth,clip]{./see_Co2MnAl2.eps}
	\caption{Seebeck coefficient of the Pt-Co$_{2}$MnAl-Pt system with 
five monolayers of Heusler between the platinum leads.} 
\label{see_Co2MnAl2}
\end{figure}
%%%%%%%%%%%%

%%%%%%%%%%%%%%%%%
\begin{figure}[t]
	\includegraphics[width=0.93\columnwidth,clip]{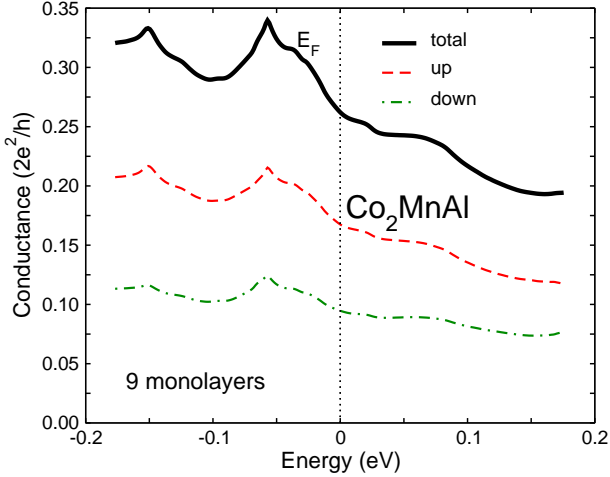}
	\caption{Energy dependence of the conductance of the 
Pt-Co$_{2}$MnAl-Pt system with nine monolayers of Heusler around the 
Fermi energy.} 
\label{conduct_Co2MnAl2}
\end{figure}
%%%%%%%%%%%%

In the system containing nine monolayers of Co$_{2}$MnAl between the platinum 
leads, the total Seebeck coefficient increases. This is contrary to the Co$_{2}$FeZ
systems where the Seebeck coefficient is smaller in the nine layer case. But here there is a sizable 
slope at low temperatures and seems to saturate for larger temperatures. 
This saturation stems from the flat regions of the energy dependence of the 
conductivity at more than 1.5 eV away from the Fermi energy. Although there
is a pronounced structure below the Fermi energy the average slope in this region
is small and therefore this region gives almost no contribution to the Seebeck
coefficient. This system is a perfect example of a system with small conductivity 
(speaking in terms of ballistic conductivity at the Femi level) but with a
large Seebeck coefficient. This shows again that the Seebeck coefficient
depends almost exclusively on the slope of the conductivity and much lesser on its absolute
value.

%%%%%%%%%%%%%%%%%
\begin{figure}[t]
	\includegraphics[width=0.93\columnwidth,clip]{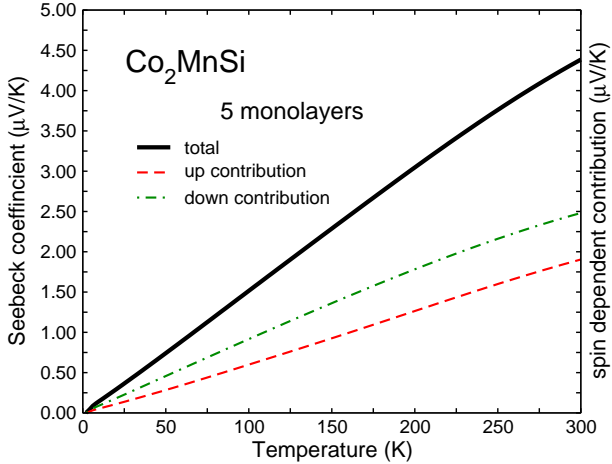}
	\caption{Seebeck coefficient of the Pt-Co$_{2}$MnSi-Pt system with 
five monolayers of Heusler between the platinum leads.} 
\label{see_Co2MnSi}
\end{figure}
%%%%%%%%%%%%

%%%%%%%%%%%%%%%%%
\begin{figure}[t]
	\includegraphics[width=0.93\columnwidth,clip]{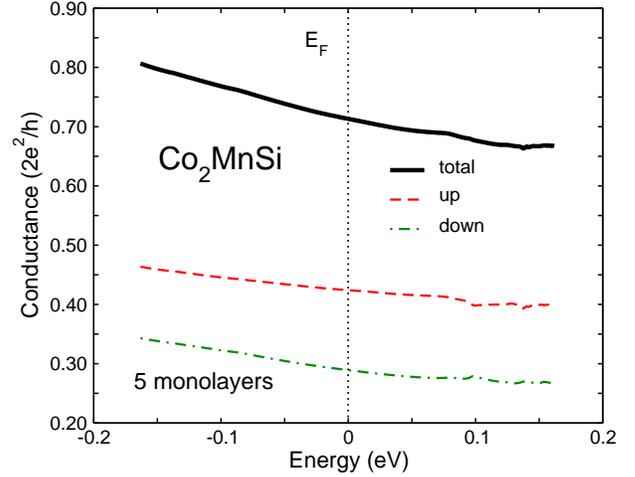}
	\caption{Energy dependence of the conductance of the 
Pt-Co$_{2}$MnSi-Pt system with five monolayers of Heusler around the 
Fermi energy.} 
\label{conduct_Co2MnSi}
\end{figure}
%%%%%%%%%%%%

To finish the discussion of the transport properties the Seebeck coefficient
of Pt-Co$_{2}$MnSi-Pt systems has to be
discussed. Figure \ref{see_Co2MnSi} shows its temperature dependence
for the system containing five monolayers Heusler. The absolute value is
decreased compared to the Pt-Co$_{2}$MnAl-Pt system. This is analogous
to the decrease found in the Co-Fe based system where the exchange of Al
by Si reduces the Seebeck coefficient strongly. The evolution of the additive
contributions from the spin channels to the total Seebeck coefficient with
temperature for Pt-Co$_{2}$MnSi-Pt is qualitatively comparable with the
one of Pt-Co$_{2}$MnAl-Pt because they are almost of the same size.
The energy dependence of the conductance shown in Fig.\,\ref{conduct_Co2MnSi}
is linear with almost no structure.

\begin{figure}[t]
	\includegraphics[width=0.93\columnwidth,clip]{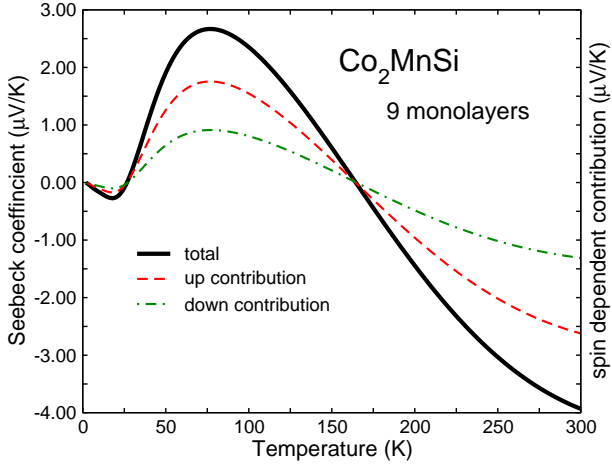}
	\caption{Seebeck coefficient of the Pt-Co$_{2}$MnSi-Pt system with 
nine monolayers of Heusler between the platinum leads.} 
\label{see_Co2MnSi2}
\end{figure}

The last system that has to be discussed is the system which contain nine monolayers
of Co$_{2}$MnSi and its Seebeck coefficient is shown in Fig.\,\ref{see_Co2MnSi2}.
Obviously this system exhibit a very interesting behavior because the Seebeck
coefficient is negative for temperatures below 25K, becomes positive up to temperature
below 175K and is again negative for higher temperatures.

%%%%%%%%%%%%%%%%%
%%%%%%%%%%%%

%%%%%%%%%%%%%%%%%
\begin{figure}[t]
	\includegraphics[width=0.93\columnwidth,clip]{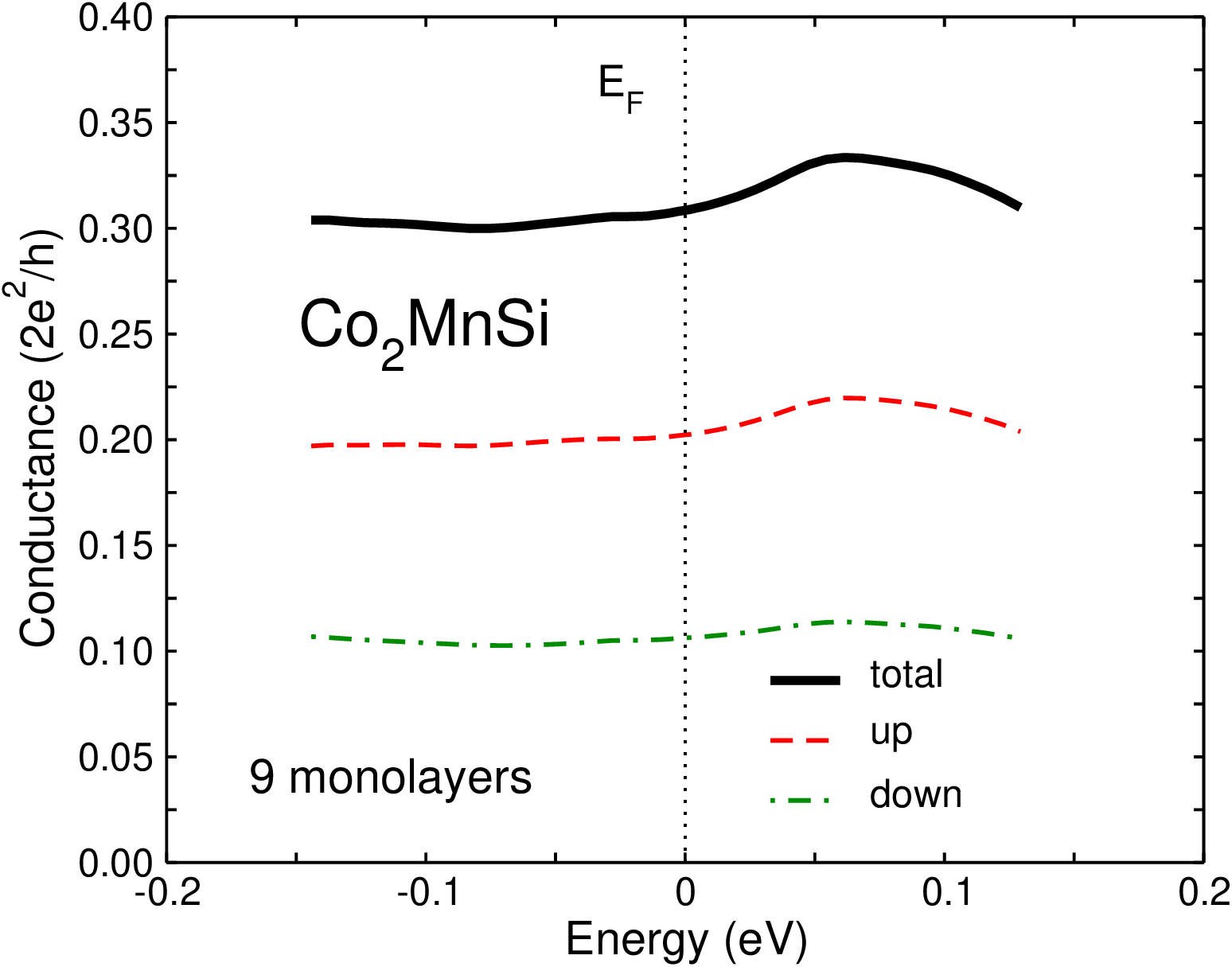}
	\caption{Energy dependence of the conductance of the 
Pt-Co$_{2}$MnSi-Pt system with nine monolayers of Heusler around the 
Fermi energy.} 
\label{conduct_Co2MnSi2}
\end{figure}
%%%%%%%%%%%%

The energy dependence of the conductance shows a bump above the energy and
a flat behavior below. This asymmetry leads to the unusual temperature dependence
of the Seebeck coefficient.

%%%%%%%%%%%%%%
\begin{table}
\begin{tabular}{ccccc}
	\hline\hline \\[-0.3cm]
	& \multicolumn{2}{c}{Al}      & \multicolumn{2}{c}{Si}    \\[-0.05cm]
%	\hline\\[-0.4cm]
	&\tiny{5ML}   &\tiny{9ML}   &\tiny{5ML}   &\tiny{9ML}\\[0.03cm]
	\hline\\[-0.3cm]
	 Fe&        + &  +       & +& -       \\ [0.1cm]
	  Mn&      + & +        & + &-       \\ [0.05cm]
	 \hline\hline
\end{tabular}
\caption{This table summarizes which combination of elements and layer thickness leads to positive
	or negative Seebeck coefficients.} 
\label{tab:plusminus}
\end{table}
%%%%%%%%%%%%

Table\,\ref{tab:plusminus} summarizes the sign of the Seebeck coefficient of all systems studied here.
It show a systematic difference between systems containing Al and those containing Si. All systems containing Al
considered here exhibit a positive Seebeck coefficient for both layer thicknesses in combination with Fe
and also with Mn. The systems that contain Si show a positive Seebeck coefficient for 5 monolayers of Heusler
for the case of Fe and Mn and a negative Seebeck coefficient for 9 monolayers in both combinations with Fe and Mn.

%%%%%%%%%%%%%%%%%%%%%%%%%%%%%%%%%%%%%%%%%%%%%%%%%%%%%%%%%%
\subsection{Electronic structure of the transport systems}
\label{sec:ElectronicStructureOfTheTransportSystems}
%%%%%%%%%%%%%%%%%%%%%%%%%%%%%%%%%%%%%%%%%%%%%%%%%%%%%%%%%%%

In order to gain a deeper insight into how the thermoelectric properties
depend on the layer thickness and the composition, the electronic structure 
of the systems discussed here have to be understood in more detail. Therefore, 
this subsection is devoted to the discussion of the electronic DOS of Heusler
layers between platinum leads. The main question is if there are signatures
of half-metallicity in the small Heusler layers and how do they influence the
Seebeck coefficient and its spin dependence.

\begin{figure}[t]
	\includegraphics[width=0.93\columnwidth,clip]{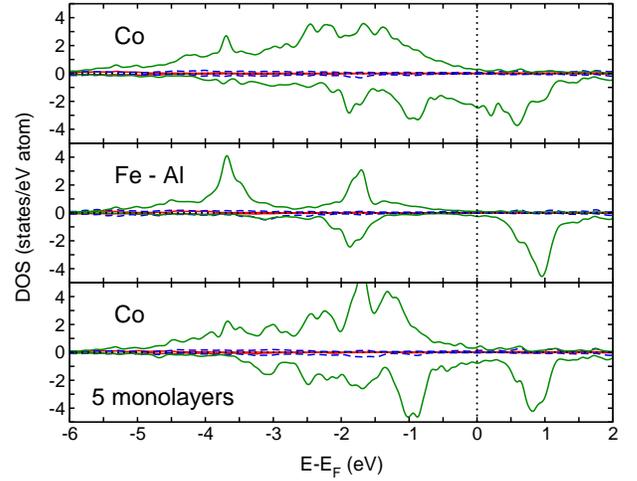}
	\caption{DOS of the Pt-Co$_{2}$FeAl-Pt system with 
five monolayers of Heusler between the platinum leads.}
\label{Co2FeAl_layer}
\end{figure}

\begin{figure}[t]
	\includegraphics[width=0.93\columnwidth,clip]{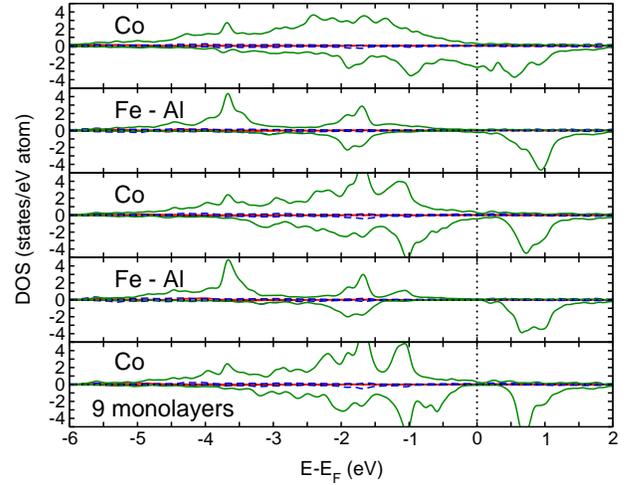}
	\caption{DOS of the Pt-Co$_{2}$FeAl-Pt system with 
nine monolayers of Heusler between the platinum leads.}
\label{Co2FeAl2_layer}
\end{figure}

In Fig.\,\ref{Co2FeAl_layer} and \ref{Co2FeAl2_layer} the electronic DOS of the two Pt-Co$_{2}$FeAl-Pt
systems are shown. The DOS of only the first three and five Heusler layers is
presented because the subsequent layers reveal the same DOS because
of the reflection symmetry of the system. One easily observes that in the system which
contains only five monolayers of Heusler the half-metallic gap is absent even
in the Co layer in the middle. But the DOS becomes obviously more similar
to that of bulk Co$_{2}$FeAl in the middle of the Heusler layer compared to the
DOS in the monolayer which is directly connected to the Pt lead.

%%%%%%%%%%%%%%%%%
%%%%%%%%%%%%

%%%%%%%%%%%%%%%%%
%%%%%%%%%%%%

If the DOS of the system that contains nine monolayers of Co$_{2}$FeAl is
examined one observes that the half-metallic gap is recovered in the middle
of the system. This means that the influence of the Pt interface is almost completely
decayed after four layers. The occurrence of this gap is responsible for certain differences
of the Seebeck coefficient between the system with five and nine monolayers.

The shape of the energy dependence of the conductance around the Fermi level
can be related to features of the DOS. Comparing the energy dependent conductances in Fig.\,\ref{conduct_Co2FeAl}
and \ref{conduct_Co2FeAl2} to the DOS in Fig.\,\ref{Co2FeAl_layer} and \ref{Co2FeAl_layer} it is
obvious why the total conductance becomes smaller from five to nine layers and
why the contribution of the two spin channel are different. For five layers both spin channel
give almost the same contribution to the conductance whereas for nine layers the contribution of the
spin up channel is almost twice as large. This can be attributed to the occurrence of the
half-metallic gap in the spin down DOS of the system with nine Heusler layers. The
absence of states in the spin down channel in the middle of the system
reduces the transmission probability of spin down electrons significantly.
The remaining transmission can be explained by the occurrence of electrons that
flipped their spin on the way through the system and by the occurrence of spin down
electron the can tunnel through the small region where there is no spin state.

%%%%%%%%%%%%%%%%%
\begin{figure}[t]
	\includegraphics[width=0.93\columnwidth,clip]{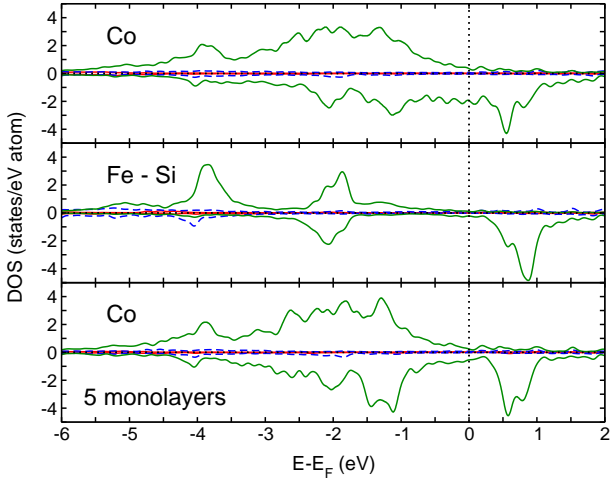}
	\caption{DOS of the Pt-Co$_{2}$FeSi-Pt system with 
five monolayers of Heusler between the platinum leads.} 
\label{Co2FeSi_layer}
\end{figure}
%%%%%%%%%%%%

%%%%%%%%%%%%%%%%%
\begin{figure}[t]
	\includegraphics[width=0.93\columnwidth,clip]{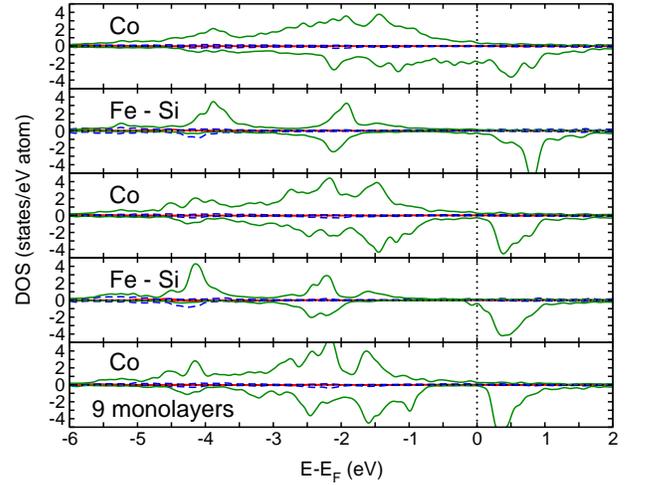}
	\caption{DOS of the Pt-Co$_{2}$FeSi-Pt system with 
nine monolayers of Heusler between the platinum leads.} 
\label{Co2FeSi2_layer}
\end{figure}
%%%%%%%%%%%%

The same arguments are also valid for the Pt-Co$_{2}$FeSi-Pt systems.
Their DOS is shown in Fig.\,\ref{Co2FeSi_layer} and \ref{Co2FeSi_layer} Again the
occurrence of the gap in the spin down channel in the nine layer system leads
to a contribution of this channel which is 50\% smaller than that of the spin up
channel. The sudden increase of the conductance above the Fermi energy in
the nine monolayer system can be related to the peak in the DOS above the Fermi
energy. The sudden occurrence of states in the middle of the system leads to the
occurrence of many new transmission channels.

\begin{figure}[t]
	\includegraphics[width=0.93\columnwidth,clip]{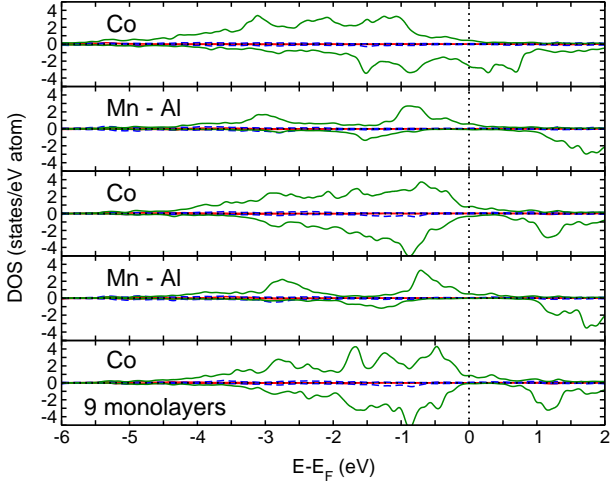}
	\caption{DOS of the Pt-Co$_{2}$MnAl-Pt system with 
nine monolayers of Heusler between the platinum leads.}
\label{Co2MnAl2_layer}
\end{figure}

The energy dependence of the five layer Pt-Co$_{2}$MnAl-Pt is also easily
described by features of the DOS which is shown in Fig.\,\ref{Co2MnAl2_layer}.
The DOS of both spin channels is relatively large
at the Fermi energy which should lead to the occurrence of many transmission
channels and therefore to a quiet large conductance. In the nine layer system
the total conductance is again reduced and also the spin down conductance is again
almost a factor of two smaller. This is again attributed to the occurrence of the
half-metallic gap (see Fig.\,\ref{Co2MnAl2_layer}).

%%%%%%%%%%%%%%%%%
\begin{figure}[t]
	\includegraphics[width=0.93\columnwidth,clip]{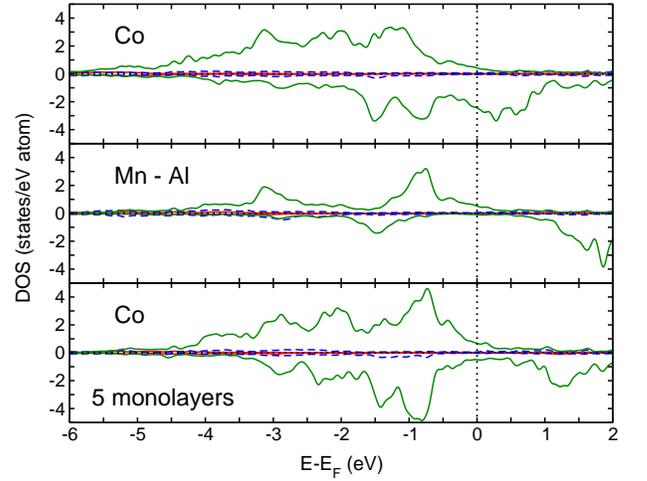}
	\caption{DOS of the Pt-Co$_{2}$MnAl-Pt system with 
five monolayers of Heusler between the platinum leads.} 
\label{Co2MnAl_layer}
\end{figure}
%%%%%%%%%%%%

%%%%%%%%%%%%%%%%%
%%%%%%%%%%%%

Concerning the DOS of the Pt-Co$_{2}$MnSi-Pt systems the results are similar
to those of the other systems. The half-metallic gap is fully recovered in the nine
monolayer system. Therefore, the small conductance contribution of the spin down channel
in this system can be attributed to this feature. Although the Fermi level of this system
is almost exactly in the middle of the gap the conductance is only small but not zero.
This shows that there must be enough tunneling channels through which electrons
can travel from one side to the other.

%%%%%%%%%%%%%%%%%
\begin{figure}[t]
	\includegraphics[width=0.93\columnwidth,clip]{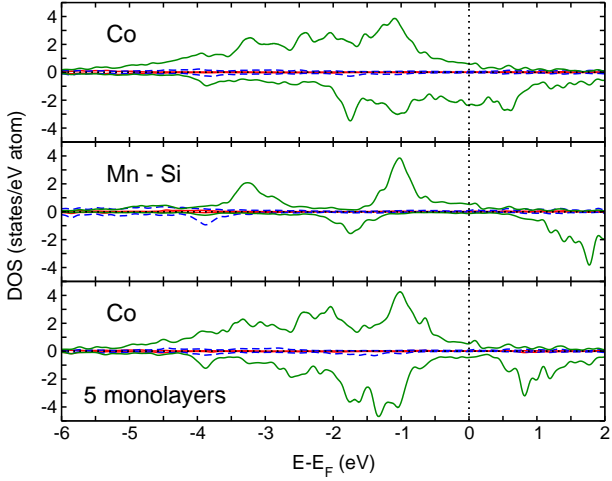}
	\caption{DOS of the Pt-Co$_{2}$MnSi-Pt system with 
five monolayers of Heusler between the platinum leads.} 
\end{figure}
%%%%%%%%%%%%

%%%%%%%%%%%%%%%%%
\begin{figure}[t]
	\includegraphics[width=0.93\columnwidth,clip]{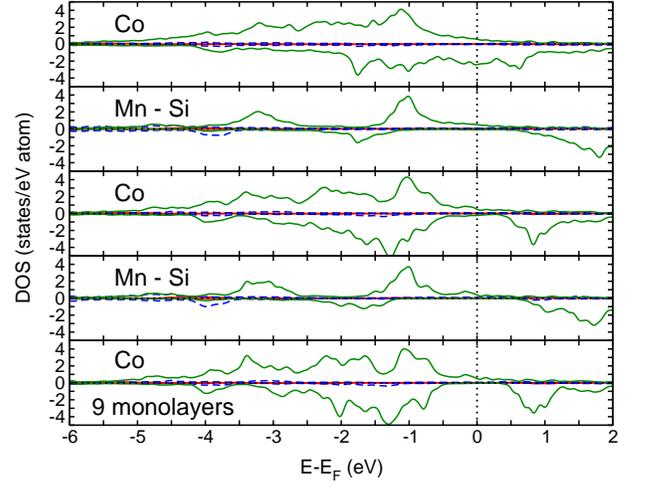}
	\caption{DOS of the Pt-Co$_{2}$MnSi-Pt system with 
nine monolayers of Heusler between the platinum leads.} 
\end{figure}
%%%%%%%%%%%%

The systematic difference between of the sign of the Seebeck coefficient as summarized in
Table\,\ref{tab:plusminus} can be related to the features of the electronic DOS. The DOS of the
Pt-Co$_{2}$FeAl-Pt and Pt-Co$_{2}$MnAl-Pt systems shows that the Fermi energy has the tendency to be located
closer to the valence edge of the spin down channel. This is different in Pt-Co$_{2}$FeSi-Pt and Pt-Co$_{2}$MnSi-Pt
where the Fermi energy shows a tendency for the conduction edge of the spin down channel.

%%%%%%%%%%%%%%%%%%%%%%%%%%%%%%%%%%%%%%%%
\section{Conclusions}\label{sec:Summary}
%%%%%%%%%%%%%%%%%%%%%%%%%%%%%%%%%%%%%%%%

Theoretical predictions of electronic, magnetic and thermoelectric properties of 
half-metallic Heusler alloys have been reported. The half-metallic shape
of the DOS of the bulk material is reproduced very nicely except for the case of Co$_{2}$FeSi.

The calculation of magnetic exchange parameters leads to an accurate reproduction
of the experimental trends of the Curie temperatures of Co-based Heuslers.
The Heisenberg description of the finite temperature magnetism of the Co-based
Heusler alloys is a good choice because the expected thermal fluctuations of
the important magnetic moments of Co, Mn and Fe are basically transversal and
therefore captured by the MC simulation of the Heisenberg model.

The transport calculations presented here are carried out for the ballistic regime.
Therefore, no inelastic scattering of electrons is considered. The temperature dependence
enters the calculation only the derivative of the Fermi function. Therewith, the additional
activation of transport channels with increasing temperature is taken into account.
Effects of phonos and magnetic excitations are neglected.

The transport calculations show that the Seebeck coefficient strongly depends 
on the details of the system. Therefore small changes of the layer thickness 
and of the composition can result in strong changes of the behavior of the 
Seebeck coefficient.

It can be stated that the Seebeck coefficient does only depend on
coarse properties of the energy dependence of the conductance. The main contribution
is given by the averaged slope of this property.

If the calculated Seebeck coefficients of the composite Pt-Heusler-Pt systems are compared to the
experimental measurements of bulk Seebeck coefficients of Co-based Heusler alloys (see Ref.\,\cite{Balke10})
fundamental differences are observed. The most obvious point is that all except the nine layer Co$_{2}$FeSi and Co$_{2}$MnSi system
show positive Seebeck coefficients over the whole temperature range. The experimental observation
of Seebeck coefficients of the corresponding bulk materials reveal negative values for all systems. It is even more surprising
that the composite systems exhibit positive Seebeck coefficients because
Pt shows also a negative Seebeck coefficient above 200K.

It clearly turns out that thin films of half-metallic Heusler alloy between platinum leads give rise to strong spin-polarized
currents and in addition to possible spin-polarized thermoelectric currents. They only need to consist of nine monolayers because
from nine monolayers on half-metallicity is recovered within the Heusler film.

The only drawback of this interpretation is that fact that the half-metallicity of Co-based Heusler alloys is strongly temperature
dependent. Therefore, the spin-polarization in such Heusler alloy is strongly reduced at higher temperatures.

%%%%%%%%%%%%%%%%
\acknowledgments
%%%%%%%%%%%%%%%%

The authors acknowledge discussions with members of the group of Hubert Ebert in particular Sebastian Wimmer and Diemo K\"{o}dderitzsch.
In addition fruitful discussion with M.\ Siewert, A.\ Gr\"{u}nebohm and M.\ E.\ Gruner and financial support within SPP1538 from DFG is acknowledged.

%%%%%%%%%%%%%%%%%%%%%%%%%%%%%%%%%%%%%%%%%%%%%%%%%%%%%%%%%%%%%%%%%%%%%%

%%%%%%%%%%%%%%%%%%%%%%%%%%%%%%%%%%%%%%%%%%%%%%%%%%%%%%%%%%%%%%%%%%%%%%

%%%%%%%%%%%%%%
\end{document}